\documentclass[12pt]{article}
																																																																																											\pdfoutput=1																																																																																			 
\usepackage{amsmath,amssymb}
\usepackage{graphicx}
\usepackage[pdftex]{hyperref}
\DeclareGraphicsExtensions{.eps,.bmp,.wmf,.jpeg,.pdf}
\numberwithin{equation}{section}
\def\be{\begin{equation}}
\def\ee{\end{equation}}

\def\bea{\begin{eqnarray}}
\def\eea{\end{eqnarray}}
\title{Higgs inflation with non-minimal derivative coupling to gravity}

\author{L. N. Granda\thanks{luis.granda@correounivalle.edu.co}\, \ D.F. Jimenez \thanks{jimenez.diego@correounivalle.edu.co}\, \ W. Cardona  \thanks{wilmar.cardona@correounivalle.edu.co}\\ {\small\it Departamento de Fisica, Universidad del Valle}\\{\small\it A.A. 25360, Cali, Colombia}}
\date{}
\begin{document}
\maketitle

\begin{abstract}
\noindent We consider an extension of Higgs inflation in which the Higgs field is non-minimally coupled to gravity through its kinetic term. We analyzed power-law coupling functions with positive or negative integer power and found that the Higgs boson can drive a successful inflation only for the cases $n=2,1,0,-1$. Theoretical predictions for both tensor to scalar ratio $r$ and scalar spectral index $n_s$ are within the 2018 \textit{Planck} $95\%$ CL. The behavior of the self coupling $\lambda$ with respect to the scalar field at the horizon crossing was obtained, and It was found that it can take values in the interval $\lambda\sim (10^{-7}, 0.3)$. 

\end{abstract}

\section{Introduction}

\noindent Several shortcomings (e.g., flatness, horizon, unobserved heavy magnetic monopoles [see for instance \cite{PhysRevLett.44.631,Linde:1987aa,Olive:1989nu,Liddle:2000cg,2009arXiv0902.1529K,Linde:1984ir}]) pervaded the standard hot big-bang scenario until around 1980 when the inflationary paradigm started to be constructed. Then, cosmologists initiated to find out that an epoch of very rapid expansion in the early universe could lead to a possible, plausible solution of the standard hot big-bang problems \cite{Starobinsky:1979ty,Starobinsky:1980te,PhysRevD.23.347,Linde:1981mu,PhysRevLett.48.1220,Linde:1983gd}. 

\noindent 
Among the wide variety of inflationary models, the simplest realization of inflation occurs with a minimally coupled scalar field \cite{Steinhardt:2001vw,Brandenberger:2008zz}. While the scalar field potential dominates the energy density of the universe, there is a period of nearly exponential growth and the drawbacks of the standard hot big-bang scenario get simply solved. Moreover, inflation also provides an explanation for the observed structures in the Universe \cite{Starobinsky:1982ee,Mukhanov:1982nu,PhysRevD.28.679,RevModPhys.57.1,Hawking:1982cz,PhysRevLett.49.1110}. Quantum fluctuations in the scalar field would have led to small inhomogeneities in the energy density that later became the structures we observe, namely, Cosmic Microwave Background (CMB) anisotropies and inhomogeneities in the mass distribution.\\
\noindent A number of approaches differing from canonical inflation have been investigated over the past years. An early inflationary regime is found in models such as non-minimally coupled scalar fields \cite{Bezrukov:2007ep,Futamase:1987ua,Fakir:1990eg}, kinetic inflation \cite{ArmendarizPicon:1999rj}, $\alpha$ attractor models \cite{Kallosh:2013yoa,PhysRevD.88.085038,Odintsov:2017yud,Dimopoulos:2017zvq}, Dirac-Born-Infeld (DBI) inflation \cite{PhysRevD.70.103505,PhysRevD.70.123505,Chen:2005ad,PhysRevD.81.023512}, string theory inspired inflation \cite{Kachru:2003sx,Kallosh:2007ig,Baumann:2014nda}, vector inflation \cite{PhysRevD.40.967,Koivisto:2008xf,Golovnev:2008cf}, inflaton potential in supergravity \cite{PhysRevLett.85.3572,Davis:2008fv,Kallosh:2010ug}, and Galileon models \cite{PhysRevD.79.064036,PhysRevD.79.084003,PhysRevD.83.083515,Ohashi:2012wf,PhysRevLett.105.231302,PhysRevD.82.103518,Burrage:2010cu,Kobayashi:2011nu},  
also including scalar fields with non-minimal derivative coupling to the Einstein tensor \cite{Capozziello:1999xt,PhysRevLett.105.011302,Granda:2011zk,articlenozari, Tsujikawa:2012mk,Yang:2015pga}.
\noindent Scalar fields are abundant in fundamental theories of matter beyond the standard model of particle physics. Nevertheless, the  only known fundamental scalar quantum field is the Higgs field discovered by the CERN collaborations ATLAS and CMS \cite{Aad:2012tfa,Chatrchyan:2012xdj}. Over the past years, intense research has been carried out trying to identify the Higgs scalar with the field causing the early inflationary period. Although this identification could in principle be possible, it has proven to be non-trivial \cite{Rubio:2018ogq}. The limitations of the semiclassical approach to slow-roll inflation with the Higgs field non-minimally coupled to gravity have been pointed out in \cite{Burgess:2009ea, barbon:2009, bezrukov:2011, calmet:2014, escriva:2017}, where the issue of unitarity bound has been addressed. The study of the unitarity issue in the new Higgs inflation with non-minimal kinetic coupling to the Einstein tensor was performed in \cite{germani:2010, germani:2010a, germani:2014, escriva:2017}, where it was shown that in this scenario there is not perturbative unitarity violation. \\
\noindent An interesting framework for inflation is provided by theories including curvature corrections such as $$\frac{\beta}{\phi^n}G_{\mu\nu}\partial^{\mu}\phi\partial^{\nu}\phi.$$ This sort of theories add gravitational friction and make the scalar field to roll slower compared to the canonical scalar field. Moreover, this sort of theories are embedded in the most general Lagrangian giving second order equations, the so-called Horndeski Lagrangian. 
There are not strongly proved physical justifications to consider higher derivative terms in the action of the scalar field to study the inflationary phenomena but, among few theoretical motivations, we can mention the fact that the type of couplings we are considering here lead to second-order field equations, avoiding the appearance of Ostrogradsky instabilities and leading to ghost-free theory, providing more general second order Lagrangian formalism. Terms of this type, with field-dependent couplings, also appear in low energy effective action of  string theory \cite{metsaev:1987, cartier:2001}. It has been shown also that couplings such as Gauss-Bonnet provide the possibility of avoiding the initial singularity \cite{antoniadis:1993, kantirizos:1999}. An appealing motivation to study such corrections to canonical scalar field is that at the high-curvature regime typical of inflation, these corrections could become appreciable and affect the outcome of inflation. 
In the case of the Higgs potential, for instance, one of the effects of the higher derivative terms is the reduction of the self coupling of the Higgs boson, so that the spectra of primordial density perturbations are consistent with the present observational data \cite{kamada:2011, ohashi:2012} \\

\noindent The main focus in this paper will be to answer to what extent the Higgs field only including a non-minimal kinetic coupling to gravity can successfully render an inflationary regime in the early universe. In section \ref{section:theoretical-framework} we explain the model and provide the theoretical framework for our study. Then we show in section \ref{section:results} that this model is compatible with current Planck constraints on the scalar spectral index $n_s$ and the tensor to scalar ratio $r$. Concluding remarks are presented in section \ref{section:conclusions}. 

\section{Theoretical framework}
\label{section:theoretical-framework}

We consider a generalization of the standard Higgs action where the Higgs field $\phi$ and its derivatives are coupled to the gravity, such that  
\be 
S = \int {{d^4}x\sqrt { - g} \left[ {\frac{R}{{2{\kappa ^2}}} - \frac{1}{2}{\partial _\mu }\phi {\partial ^\mu }\phi  - \frac{\lambda }{4}{\phi ^4} + \frac{\beta }{{{\phi ^n}}}{G_{\mu \nu }}{\partial ^\mu }\phi {\partial ^\nu }\phi } \right]},
\label{Eq:action}
\ee
where $R$ is the Ricci scalar, $\kappa^2 \equiv 8 \pi G_N$ with $G_N$ the bare Newton's constant, $G_{\mu \nu }$ is the Einstein tensor, $\beta$ is an arbitrary parameter with dimension of $(mass)^{n-2}$ and $n$ is an integer. We will further assume that during inflation one can neglect the vacuum expectation value of the Higgs scalar so that a quartic potential is a good approximation for the Higgs potential. 
The self-coupling constant $\lambda$ is fixed experimentally by the LHC results, which impose the constrain $\lambda \simeq 0.13$ at the electroweak scale \cite{PhysRevD.98.030001}. \\
In the flat Friedmann-Lemaitre-Robertson-Walker (FLRW) metric the equations of motion take the form
\be\label{friedamanneq}
H^2=\frac{\kappa^2}{3}\left(\frac{1}{2}\dot{\phi}^2+\frac{\lambda}{4}\phi^4+9\frac{\beta}{\phi^2}H^2\dot{\phi}^2\right)
\ee
\be\label{friedman2}
2\dot{H}F\left(1-\frac{\kappa^2\beta\dot{\phi}^2}{\phi^n}\right)=-\kappa^2\left(\dot{\phi}^2+6\beta \frac{H^2\dot{\phi}^2}{\phi^n}-4\beta \frac{H\dot{\phi}\ddot{\phi}}{\phi^n}+2n\beta\frac{H\dot{\phi}^3}{\phi^{n+1}}\right)
\ee
and
\be\label{fieldeq}
\ddot{\phi}+3H\dot{\phi}+\lambda\phi^3+ 6\beta H(3H^2+2\dot{H})\frac{\dot{\phi}}{\phi^2}+6\beta H^2\frac{\ddot{\phi}}{\phi^2}-6\beta H^2\frac{\dot{\phi}^2}{\phi^3}=0.
\ee
Next we define the slow-roll parameters, which involve the interaction terms in the model (\ref{Eq:action}), as  (see \cite{Granda:2019wpe})
\be\label{epsilon0}
\epsilon_0=-\frac{\dot{H}}{H^2},\;\;\; \epsilon_1=\frac{\dot{\epsilon}_0}{H\epsilon_0}
\ee
\be\label{k0}
k_0=3\beta\kappa^2\frac{\dot{\phi}^2}{\phi^n},\;\;\; k_1=\frac{\dot{k}_0}{Hk_0}
\ee
Note that for the simple canonical scalar field, $\frac{1}{2}\dot{\phi}^2+V(\phi)$, the above $H$-defined slow-roll parameters are related to the standard $V$-defined slow-roll parameters
\be\label{v-slow-roll}
\epsilon_v=\frac{M_p^2}{2}\left(\frac{V'}{V}\right)^2,\;\;\; \eta_v=M_p^2\frac{V''}{V}
\ee
in the way
\be\label{comparing-H-V}
\epsilon_0=\epsilon_v, \;\;\; and \;\;\;  \epsilon_1=4\epsilon_v-2\eta_v
\ee
where the slow-roll approximation 
$$ 2H^2\simeq \kappa^2 V,\;\;\; 3H\dot{\phi}+V'\simeq 0$$
was used.\\
Writing the cosmological equations (\ref{friedamanneq})-(\ref{fieldeq}) in terms of the slow-roll, parameters (\ref{epsilon0}), (\ref{k0}), we can express the derivative of the scalar field and the potential as
\be\label{slow-roll-V}
V=M_p^2H^2\left[3-\epsilon_0 -2k_0 -\frac{1}{3}k_0\left(k_1-\epsilon_0\right)\right]
\ee
\be\label{slow-roll-kinetic}
\dot{\phi}^2=M_p^2H^2\left[2\epsilon_0-2k_0+\frac{2}{3}k_0\left(k_1-\epsilon_0\right)\right]
\ee
from which it becomes clear that $\dot{\phi}^2<<V$ under the slow-roll conditions $\epsilon_0,k_0,...<<1$. The slow-roll conditions $\dot{\phi}^2<<V=\frac{\lambda}{4}\phi^4$, $\ddot{\phi}<<3H\dot{\phi}$, $k_0,k_1<<1$ allow to write the Eqs. (\ref{friedamanneq})-(\ref{fieldeq}) in the following approximate way
\be\label{slow-roll-H}
H^2\simeq \frac{\kappa^2}{3}V=\frac{\kappa^2}{3}\left(\frac{\lambda}{4}\phi^4\right)
\ee
\be\label{slow-roll-dotH}
\dot{H}\simeq -\frac{\kappa^2}{2}\left(\dot{\phi}^2+6\beta\frac{H^2\dot{\phi}^2}{\phi^n}\right)
\ee
\be\label{slow-roll-phi}
3H\dot{\phi}+V'+18\beta\frac{H^3\dot{\phi}}{\phi^n}\simeq 0
\ee
From these equations we can see that the potential term gives the dominant contribution to the Hubble parameter, and Eqs. (\ref{slow-roll-dotH}), ()\ref{slow-roll-phi}) determine the role of the kinetic coupling in the slow-roll dynamics.  These equations allow to find the number of $e$-folds as 
\be\label{e-folds}
N=\int_{\phi_I}^{\phi_E}\frac{H}{\dot{\phi}}d\phi=-\int_{\phi_i}^{\phi_{end}}\frac{3H^2}{V'}\left(1+6\beta\frac{H^2}{\phi^n}\right)d\phi=-\int_{\phi_i}^{\phi_{end}}\frac{V}{M_p^2V'}\left(1+2\beta\frac{V}{M_p^2\phi^n}\right)d\phi
\ee
where $\phi_i$ and $\phi_{end}$ are the values of the scalar field at the beginning and end of inflation respectively. The scalar field at the end of inflation can be found from the condition $\epsilon_0(\phi_{end})=1$, and $\phi_i$, which is the scalar field at the horizon crossing, is calculated from (\ref{e-folds}) by giving the appropriate number of e-folds such that the  inflationary observables take values consistent with current observations. In our case we observed that given $N\simeq 60$ produce the necessary inflation to set the values of the scalar spectral index and the tensor-to-scalar ratio in the most appropriate regions, within the capabilities of the model. For the $\phi^4$ potential we find 
\be
N \approx \int\limits_{{\phi _i}}^{{\phi _{end}}} {\frac{1}{4}\phi \left( {\frac{1}{2}\beta\lambda {\phi ^{4 - n}} + 1} \right)d\phi }, 
\label{nfolds}
\ee 
In order to obtain the inflationary observable magnitudes we need to calculate the power spectrum of the scalar and tensor perturbations, which was presented for  a general class scalar-tensor models in  \cite{Granda:2019wpe}. For our particular case, as follows from  \cite{Granda:2019wpe}, the scalar spectral index and the tensor-to-scalar ratio are given by (in first order in slow-roll parameters)  
\be\label{spectral-phi-h}
n_s=1-2\epsilon_0-\epsilon_1,\;\;\;  r=16\epsilon_0
\ee
where we can see that the non-minimal kinetic coupling to curvature does not affect the standard consistency relation, at least up to first order in slow-roll parameters.
These magnitudes can be written as functions of the scalar field and are evaluated at the horizon crossing. 
To this end we express the Hubble parameter and its derivatives, $\dot{H}$ and $\ddot{H}$, together with the derivatives of the scalar field $\dot{\phi}$, $\ddot{\phi}$ in terms of the potential and the kinetic coupling function, reducing all magnitudes to functions of the scalar field. We can illustrate this procedure for the general case of model with potential $V(\phi)$ and kinetic coupling $F_1(\phi)$, writing the Eqs. (\ref{slow-roll-H})-(\ref{slow-roll-phi}) in the form
\be\label{slow-roll-H1}
H^2\simeq \frac{\kappa^2}{3}V(\phi)
\ee
\be\label{slow-roll-dotH1}
\dot{H}\simeq -\frac{\kappa^2}{2}\left(\dot{\phi}^2+6H^2F_1(\phi)\dot{\phi}^2\right)
\ee
\be\label{slow-roll-phi1}
3H\dot{\phi}+V'+18H^3F_1(\phi)\dot{\phi}\simeq 0.
\ee
The Eq. (\ref{slow-roll-H1}) gives directly $H(\phi)$, while combining the three equations and taking the necessary derivatives we find
\be\label{dot-H}
\dot{H}=-\frac{V'(\phi)^2}{6V(\phi)\left(1+2\kappa^2 F_1(\phi) V(\phi)\right)}
\ee
\be\label{dot-phi}
\dot{\phi}=\frac{V'(\phi)}{\sqrt{3}\kappa\sqrt{V}\left(1+2\kappa^2 F_1(\phi) V(\phi)\right)}
\ee
\be\label{ddot-H}
\ddot{H}=\frac{V'(\phi)^2\left[V'(\phi)^2+V\left(4\kappa^2F_1(\phi)V'(\phi)^2-2V''(\phi)\right)+2\kappa^2 V(\phi)^2\left(F_1'(\phi)V'(\phi)-2F_1(\phi) V''(\phi)\right)\right]}{6\sqrt{3}\kappa V(\phi)^{5/2}\left(1+2\kappa^2 F_1(\phi) V(\phi)\right)^3}
\ee
\be\label{ddot-phi}
\ddot{\phi}=-\frac{V'(\phi)\left[V'(\phi)^2+V(\phi)\left(6\kappa^2F_1(\phi) V'(\phi)^2-2V''(\phi)\right)+4\kappa^2V(\phi)^2\left(F_1'(\phi) V'(\phi)-F_1(\phi) V''(\phi)\right)\right]}{6\kappa^2 V(\phi)^2\left(1+2\kappa^2 F_1(\phi) V(\phi)\right)^3}
\ee
Replacing these results in (\ref{epsilon0}) and (\ref{k0}), and turning to the specific case $V=\lambda\phi^4/4$ and $F_1=\beta/\phi^n$, after some algebra we find 
\be 
{\epsilon_0} = \frac{{16}}{{2{\phi ^2} + \alpha {\phi ^{6 - n}}}}, \;\;\; \epsilon_1=\frac{\left[32\phi^n+8(6-n)\alpha\phi^4\right]\phi^{n-2}}{\left(2\phi^n+\alpha\phi^4\right)^2}
\label{Eq:slow-roll-parameter-0}
\ee
\be\label{k0-k1}
k_0=\frac{16\alpha\phi^{n+2}}{\left(2\phi^n+\alpha\phi^4\right)^2},\;\;\; k_1=\frac{\left[16(n-2)\phi^n+8(6-n)\alpha\phi^4\right]\phi^{n-2}}{\left(2\phi^n+\alpha\phi^4\right)^2}
\ee
where
\be\label{alpha}
 \alpha\equiv \beta\lambda
\ee 
 and we have set  $M_p=1$. Finally, using (\ref{spectral-phi-h}), we can write the  scalar spectral index and the tensor-to-scalar ratio in terms of the scalar field for the model (\ref{Eq:action}) as
\be\label{Eq:scalar-spectral-index}
{n_s} = \frac{{{\alpha ^2}{\phi ^{10}} + 4\alpha \left( {2n + {\phi ^2} - 20} \right){\phi ^{n + 4}} + 4\left( {{\phi^2} - 24} \right){\phi^{2n}}}}{{{\phi^2}{{\left( {\alpha {\phi ^4} + 2{\phi^n}} \right)}^2}}}\Big |_{\phi=\phi_i},
\ee
\be\label{Eq:tensor-to-scalar-ratio}
r =\frac{{256}}{{2{\phi^2} + \alpha {\phi^{6 - n}}}}\Big |_{\phi=\phi_i}.
\ee 
In order to analyze the observational constraints on the model, and particularly, how they influence the self coupling $\lambda$, we resume here some aspects of the second order formalism (see \cite{Granda:2019wpe} for details). \\
For the scalar and tensor perturbations we can write the power spectra respectively as
\be\label{slrt13}
P_{\xi}=A_S\frac{H^2}{(2\pi)^2}\frac{{\cal G}_S^{1/2}}{{\cal F}_S^{3/2}},\;\;\; P_T= 16A_T\frac{H^2}{(2\pi)^2}\frac{{\cal G}_T^{1/2}}{{\cal F}_T^{3/2}}
\ee
where 
\be\label{velocity}
{\cal F}_S=c_S^2{\cal G}_S, \;\;\; {\cal F}_T=c_T^2{\cal G}_T
\ee
and 
\be\label{coeffa}
A_S=\frac{1}{2}2^{2\mu_s-3}\Big|\frac{\Gamma(\mu_s)}{\Gamma(3/2)}\Big|^2,\;\;\; A_T=\frac{1}{2}2^{2\mu_T-3}\Big|\frac{\Gamma(\mu_T)}{\Gamma(3/2)}\Big|^2,
\ee
where in terms of the slow-roll parameters, up to first order
\be\label{slr2}
{\cal G}_S=M_p^2 \epsilon_0
\ee
and 
\be\label{velocity1}
c_S^2=1,
\ee
 \be\label{gt}
 {\cal G}_T=M_p^2\left(1-\frac{1}{3}k_0\right)
 \ee
 and 
 \be\label{tensorvelocity}
 c_T^2=\frac{3+k_0}{3-k_0}
 \ee
where $c_S$ and $c_T$ are the velocities of scalar and tensor perturbations respectively, and all magnitudes are evaluated at the moment of horizon exit when $c_S k=aH$.\\ 
On the other hand, the relative contribution to the power spectra of tensor and scalar perturbations, defined as the tensor/scalar ratio $r$, given by
\be\label{slrt12}
r=\frac{P_T(k)}{P_{\xi}(k)},
\ee
can be written as
\be\label{slrt15}
r=16\frac{{\cal G}_T^{1/2}{\cal F}_S^{3/2}}{{\cal G}_S^{1/2}{\cal F}_T^{3/2}}=16\frac{{\cal G}_S}{{\cal G}_T}
\ee
where we used fact that $A_T/A_S\simeq 1$ and $c_T\simeq 1$ when evaluated at the limit $\epsilon_0,...<<1$. Using (\ref{slr2}) and (\ref{gt}), gives the consistency relation for $r$ (\ref{spectral-phi-h}).
The equation (\ref{slrt13}) for $P_{\xi}$ with (\ref{coeffa}) for $A$ allows to find scale of the Hubble parameter during inflation by using the COBE normalization for the power spectra, which can be written as  (\ref{slrt13}) 
\be\label{power-cobe}
P_{\xi}=A_S\frac{H^2}{(2\pi)^2}\frac{{\cal G}_S^{1/2}}{{\cal F}_S^{3/2}}\sim \frac{H^2}{2(2\pi)^2}\frac{1}{{\cal F}_S}\sim  \frac{H^2}{(2\pi)^2M_p^2}\frac{1}{2\epsilon_0}
\ee
where at the limit $(\epsilon_0,\epsilon_1,...)\rightarrow 0$ we used the approximation $A_S\rightarrow 1/2$.
According to COBE normalization $P_{\xi}\simeq 2.4\times 10^{-9}$. Then,
\be\label{cobe1}
H^2\sim 2(2.4\times 10^{-9})(2\pi)^2\epsilon_0.
\ee
On the other hand, from the tensor/scalar ratio we find
\be\label{cobe2}
P_T=rP_{\xi}\sim \frac{2V}{3\pi^2M_p^4}\;\;\;  \Rightarrow \;\;\; V\sim ( 2.4\times 10^{-9})\frac{3\pi^2}{2} r M_p^4,
\ee
which also follows from (\ref{cobe1}). For the Higgs potential this last result allows to find the self coupling $\lambda$ in terms of the scalar field at the beginning of inflation as
\be\label{lambda-phi}
\lambda\sim \frac{1536  (2.4\times 10^{-9})\pi^2}{\phi^4\left(2\phi^2+\alpha\phi^{6-n}\right)}\Bigg |_ {\phi_i}
\ee
From this result we can see that $\lambda$ decreases with the increment of the scalar field at the horizon crossing. As will be shown in some numerical cases, large field inflation can give $\lambda\sim 10^{-6}$. When $\phi_i$ takes almost constant small values, at large $\alpha$, the self coupling can take values of the order $\lambda\sim 0.1$. In the next section we consider some numerical cases.

\section{Results}
\label{section:results}

\begin{figure}
\centering
\includegraphics[scale=0.7]{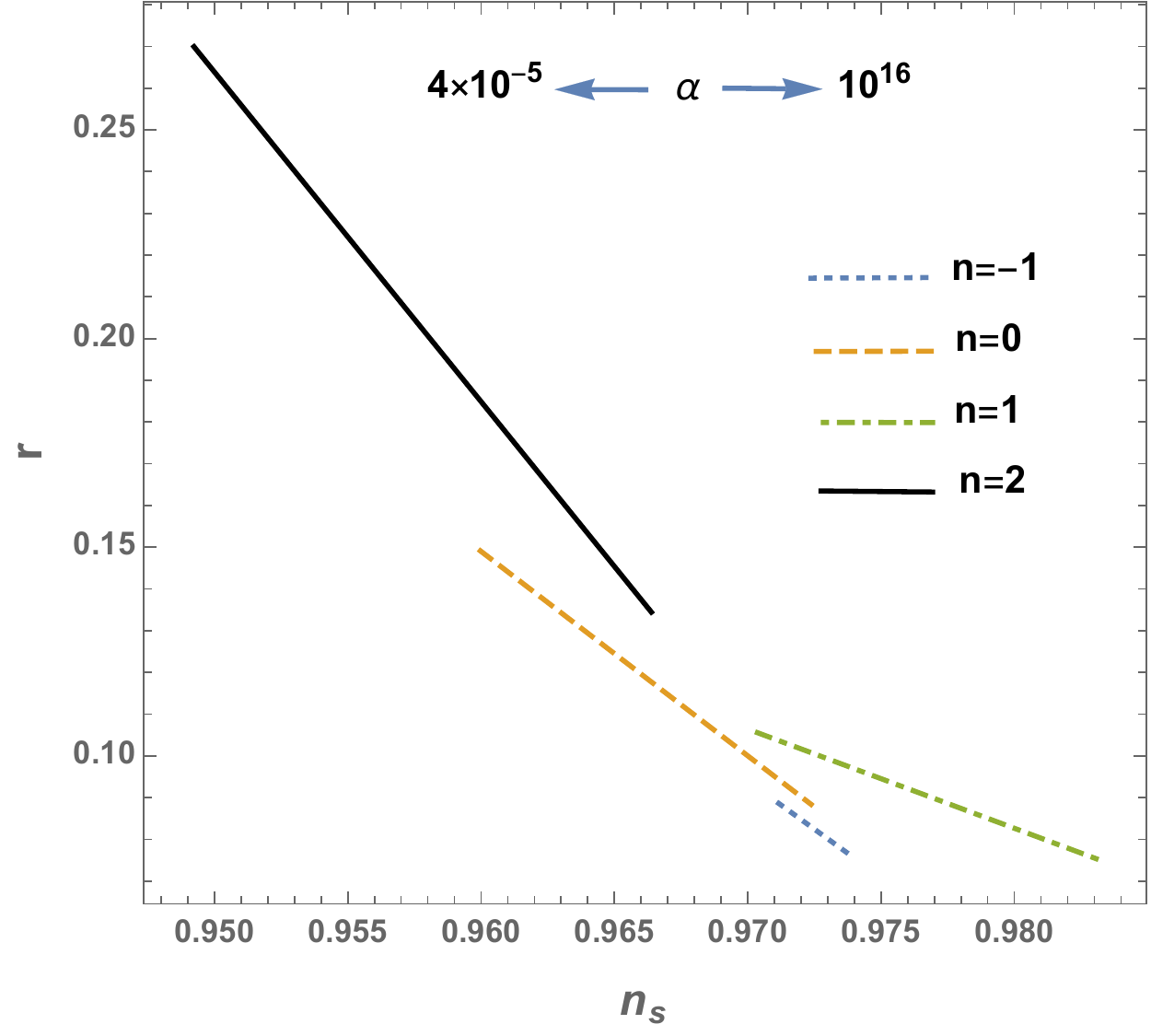}
\caption{Parametric plot of $r$ vs $n_S$ for $4 \times {10^{ - 5}} < \alpha  <  {10^{16}}$ with $N=60$, for coupling functions corresponding to $n=-1,\,0,\,1,\,2$. For the models $n=-1$ and $n=0$ the inflationary observables $n_s$ and $r$ are more close to the region favored by current observational data.}
\label{rvsnskin}
\end{figure}

\begin{figure}
\centering
\includegraphics[scale=0.55]{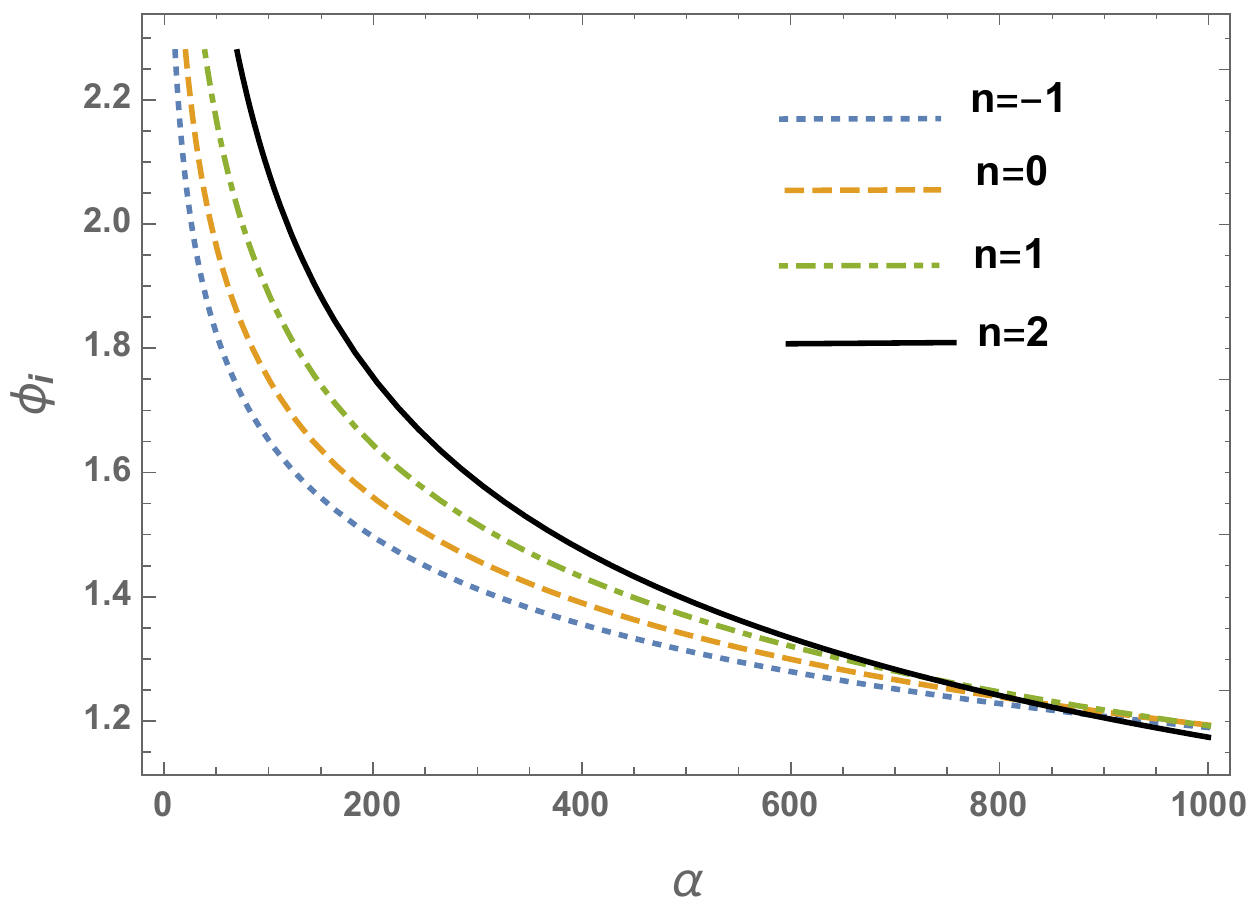}
\includegraphics[scale=0.55]{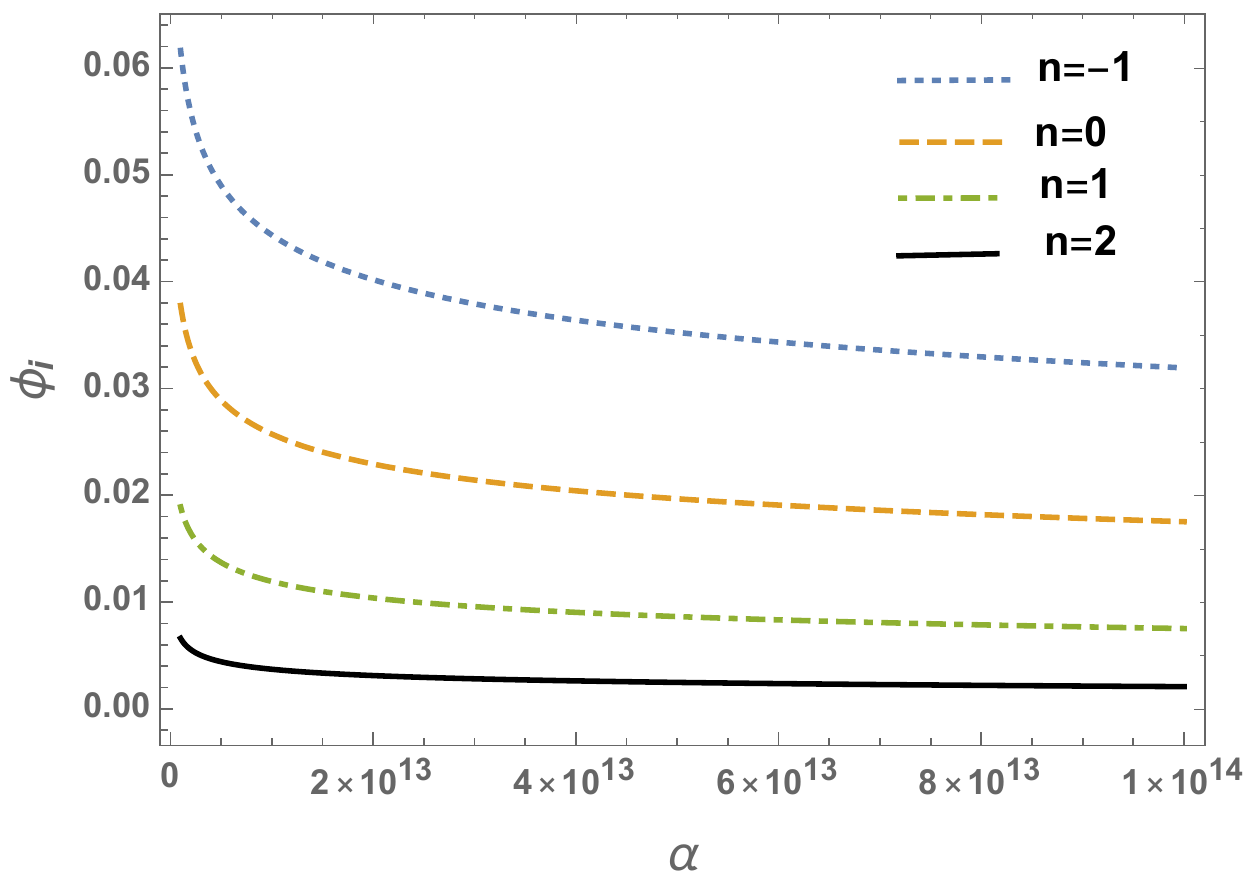}
\caption{The inflaton at the horizon crossing as function of the parameter $\alpha$ for $n=-1,0,1,2$ with $N=60$. The left graphic shows large field inflation scenario where, for $\alpha\lesssim 10^{3}$, the scalar filed takes values $\phi_1\gtrsim M_p$. The right graphic shows the small field inflation, where starting from $\alpha\gtrsim 10^{13}$, the initial scalar field becomes $\phi_{i}\lesssim 10^{-2}$ $60$-$e$-folds before the end of inflation and continue decreasing with the increase of $\alpha$.}
\label{phiini}
\end{figure}

\begin{figure}
\centering
\includegraphics[scale=0.7]{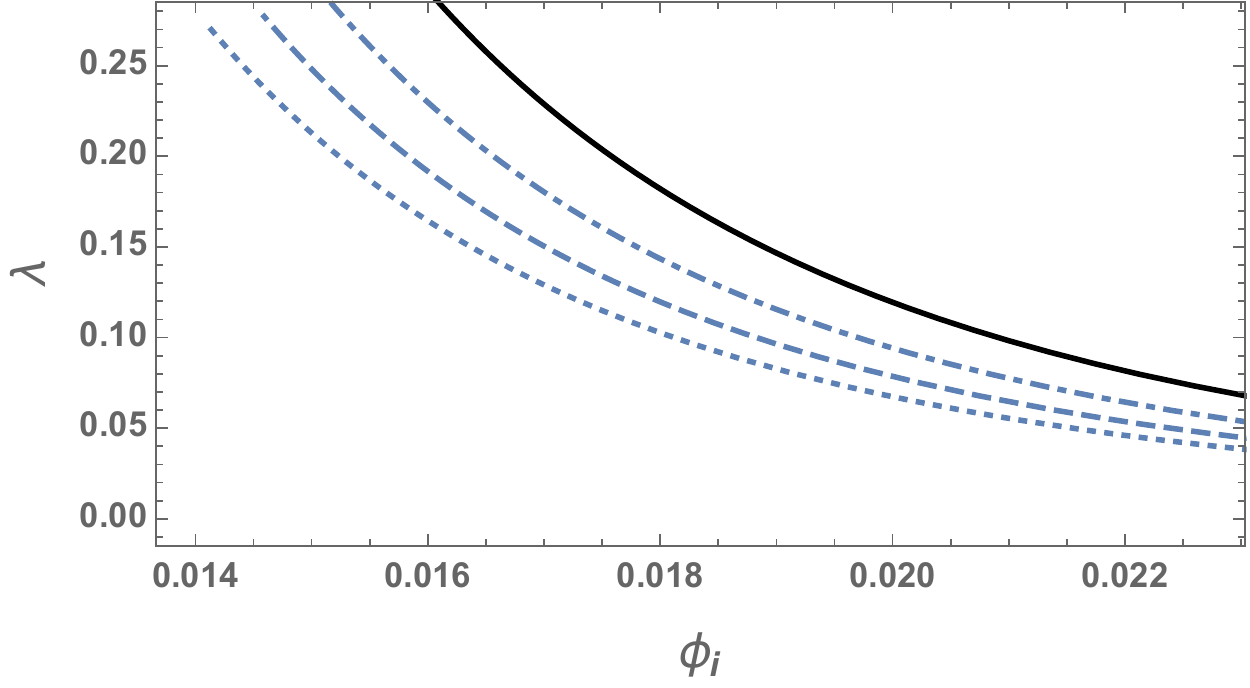}
\caption{The behavior of the self coupling $\lambda$ with respect to the scalar field at the horizon crossing, $60$ $e$-folds before the end of inflation, for the four models corresponding to $n=-1,0,1,2$. These curves correspond to the small-field inflation, giving $\lambda\sim 10^{-2}-0.28$ .}
\label{lambda-phi1}
\end{figure}

\begin{figure}
\centering
\includegraphics[scale=0.75]{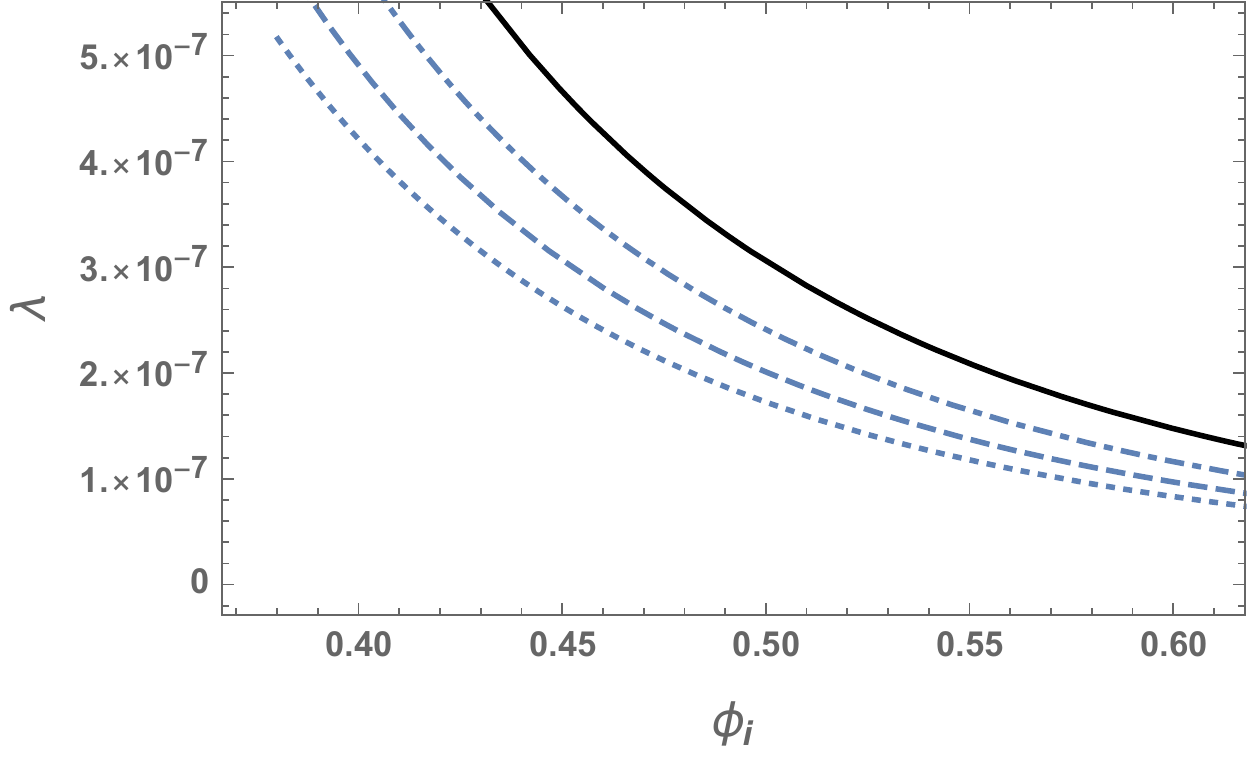}
\caption{The self coupling $\lambda$ for a large-field ($\phi\sim M_p$) inflationary scenario corresponding to the left graphic in Fig. (\ref{phiini}). Along the shown interval $\lambda$ remains of the order of $10^{-7}$.}
\label{lambda-phi2}
\end{figure}

\noindent In this section we show theoretical predictions of the model for both tensor to scalar ratio $r$ and scalar spectral index $n_s$. In Fig. \ref{rvsnskin} we consider four coupling functions corresponding to $n=-1,\,0,\,1,\,2$ and show the behavior of $r$ and $n_s$ for a wide $\alpha$-interval. Since the field at horizon crossing depends on $\alpha$, it is therefore important to make sure that the inflationary period occurs for field values such that the current restrictions on the self coupling $\lambda$ of the Higgs boson are satisfied.  
In Fig. 2  we show the behavior of the inflaton field at the horizon crossing,$60$ $e$-folds before the end of inflation, with respect to $\alpha$. The left graphic shows that in the region $\alpha\lesssim 10^{3}$, the scalar filed takes values $\phi_i\gtrsim M_p$ leading to large field inflation, and the right graphic shows that in the region $\alpha\sim 10^{12}-10^{14}$ the scalar field takes values of the order $\phi_i\sim 10^{-2}$. This affects the behavior of the self coupling $\lambda$ as shown in Figs. 3 an 4 below, where $\lambda$ can take values in the interval $10^{-7}\lesssim\lambda\lesssim 0.3$, covering a wide spectrum of possible values of the self coupling, according to the different Higgs boson phenomenological restrictions.
\noindent In Fig. \ref{phiini} we show the variation of the scalar field at horizon crossing. Notice that at large values ($\alpha \sim 10^{13}-10^{14}$) the field $\phi_{i} \sim 10^{-2}M_p$ and at small $\alpha$, $\alpha\lesssim 10^3$ one finds $\phi_1\gtrsim M_p$. \\
In Figs. \ref{lambda-phi1},  \ref{lambda-phi2} we show the behavior of the self coupling $\lambda$ with respect to the scalar field at the horizon crossing, where due to the difference in the scale we show two different intervals for $\phi_i$.
It is important to note that along both $\alpha$-intervals, which results in the corresponding intervals for the scalar field in Figs. 3 and 4, the scalar spectral index $n_s$ and tensor/scalar ratio $r$ remain practically unchanged for each case, taking the following values  
$$ n=-1\;\;\; \Rightarrow \;\;\;   n_s\simeq 0.974, \; r\simeq 0.075$$
$$ n=0 \;\;\; \Rightarrow \;\;\;   n_s\simeq 0.972, \; r\simeq 0.088$$
$$ n=1 \;\;\; \Rightarrow \;\;\;   n_s\simeq 0.970, \; r\simeq 0.106$$
$$n=2 \;\;\; \Rightarrow \;\;\;   n_s\simeq 0.966, \; r\simeq 0.134$$
the difference between the values at $\alpha\sim 10^3$ and $\alpha\sim 10^{16}$ is in the order of $10^{-4}$ or less, while changes a little more for $\alpha\lesssim 10^{-3}$ where the difference can be of the order of $10^{-2}-10^{-1}$ (which is more accentuated for $r$), getting closer to the weak coupling limit case.\\
\noindent Furthermore, it is worth noticing that the contribution of the kinetic term is subdominant in comparison with the potential. This can be easily seen if we take $\beta\sim10^{15}$, and $N=60$ which after applying the COBE normalization gives $H\sim 3.2\times 10^{-5}M_p$. Taking for instance the case $n=0$ and computing the kinetic coupling term one obtains the following value 
\[9\beta{H^2}\dot \phi _c^2 = 9\beta{H^4}{\left( {\frac{{d{\phi _c}}}{{dN}}} \right)^2} \simeq 2.2 \times {10^{ - 11}}M_p^4 \,\,\,\,\,\,\,\,\,  \]
where $d\phi_{c}/dN$ is obtained by taking the derivative of Eq. (\ref{nfolds}) with respect to the e-foldings number. It becomes clear that the kinetic term is much smaller than the potential, which for the above cases is of order of $V(\phi_c) \sim{10^{-9}}M_p^4$ (after COBE normalization), so that the inflationary period is driven mainly by the Higgs potential. 

\begin{figure}[hbtp]
\centering
\includegraphics[scale=0.75]{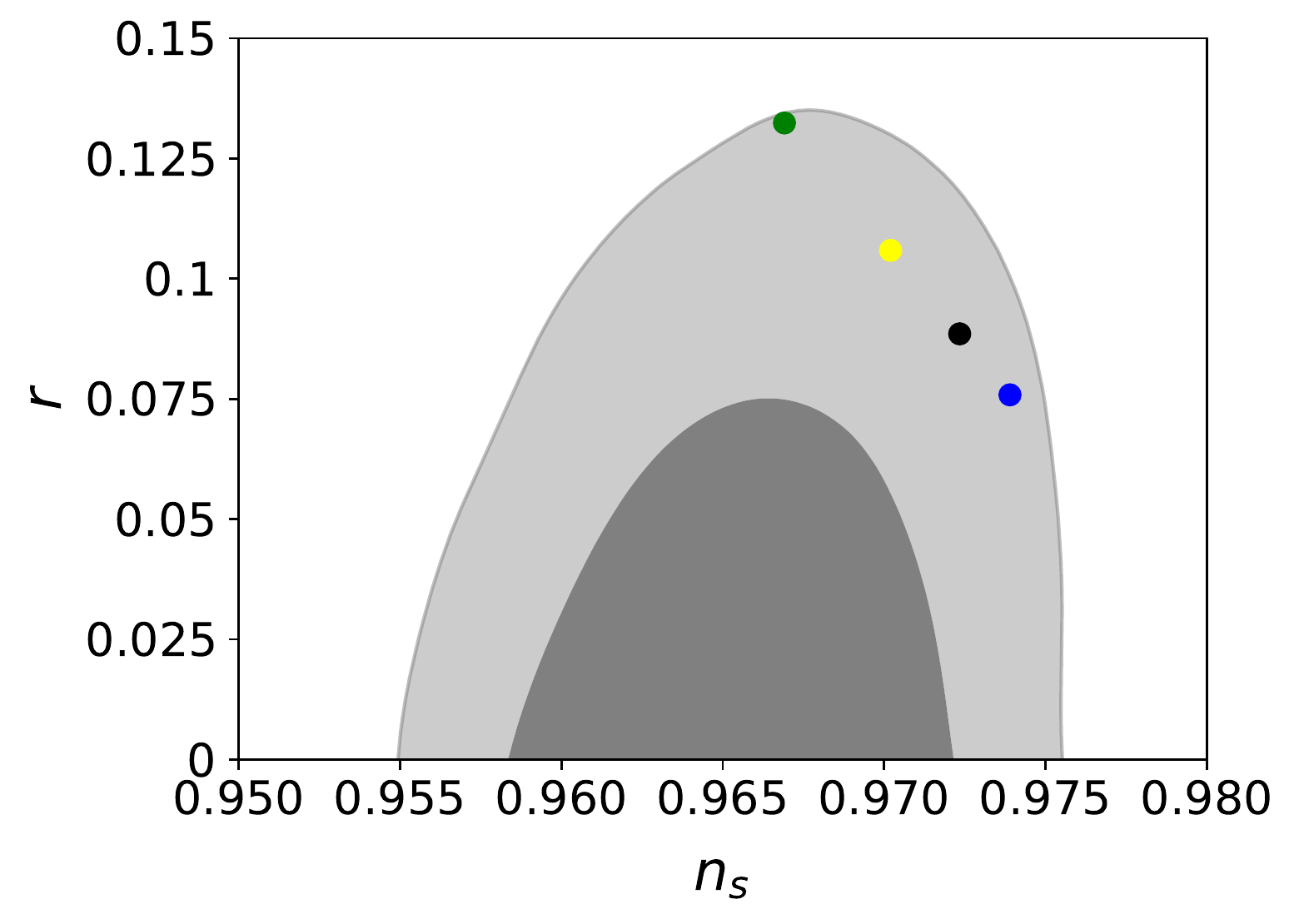}
\caption{Marginalized $68\%$ and $95\%$ CL regions for the scalar spectral index $n_s$ and the tensor to scalar ratio $r$ at $k=0.0002\, \rm{Mpc}^{-1}$. We use the publicly available chains   including 2018 Planck data alone, namely, \texttt{plikHM\_TTTEEE\_lowl\_lowE\_lensing}. Colored dots show theoretical predictions for the models $\beta/\phi^2$ (green), $\beta/\phi$ (yellow), $\beta$ (black), and $\beta \phi$ (blue). These predictions consider $N=60$ and $\alpha\sim 10^{14}$.}
\label{Fig:planck-results}
\end{figure}

\noindent In Fig. \ref{Fig:planck-results} we show the theoretical predictions of the model for $\alpha=10^{14}$ embedded in the 2018 \textit{Planck} constraints including temperature, polarization, and lensing. It is shown the $68\%$ and $95\%$ CL regions for the scalar spectral index $n_s$ and the tensor to scalar ratio $r$. Colored dots correspond to the theoretical predictions for the models $\beta/\phi^2$ (green), $\beta/\phi$ (yellow), $\beta$ (black), and $\beta \phi$ (blue). One can see that for all the cases the theoretical predictions fall inside the $95\%$ CL region. Moreover, the effect of the non-minimal kinetic coupling is more pronounced in the scalar spectral index than in the tensor to scalar ratio. It is worth noticing that considering $N=50$ $e$-folds the tensor-to-scalar ratio increments for all cases, while at $N=70$ the increment is on the scalar spectral index side, making in both cases the models less viable.

\noindent Analyzing the model in the strong coupling limit we can extract some conclusions regarding the behavior of the observables $n_s$ and $r$ for different integer powers $n$. In absence of kinetic coupling all models reduce to the standard canonical field with $\phi^4$ potential that gives the know values for $n_s$ and $r$ given by
\be\label{weaklimit}
n_s=\frac{N-2}{N+1},\;\;\; r=\frac{16}{N+1},
\ee
and at the strong coupling limit ($\beta\rightarrow \infty$) we find the following values
\be\nonumber
n_s^{(2)}=\frac{2N-3}{2N+1},\;\;\; r^{(2)}=\frac{16}{2N+1},\;\;\; n_s^{(1)}=\frac{5N-7}{5N+2},\;\;\; r^{(1)}=\frac{32}{5N+2},
\ee
\be\label{stronglimit}
n_s^{(0)}=\frac{3N-4}{3N+1},\;\; r^{(0)}=\frac{16}{3N+1},\;\; n_s^{(-1)}=\frac{7N-9}{7N+2},\;\; r^{(-1)}=\frac{32}{7N+2},
\ee
where the upper labels correspond to the power $n$. These values of $r$ at the strong coupling limit correspond to the minimum value that $r$ can achieve for the corresponding model. Assuming $N=60$ we find that the models with the constant coupling $\beta$ ($n=0$) and $\beta\phi$ ($n=-1$) give the lower tensor-to-scalar ratio, $ r^{(0)}\approx 0.088$ and $r^{(-1)}\approx 0.076$, that are below $0.1$. For $n=3$ at the strong coupling limit the observables take the values
$$
n_s^{(3)}=\frac{3N-5}{3N+2},\;\; r^{(3)}=\frac{32}{3N+2},
$$
which at $N=60$ gives an appropriate value $n_s\approx 0.961$ but large tensor-to-scalar ratio $r\approx 0.176$. 
If we take  $n=4$, then $n_s$ and $r$ remain the same as in the absence of coupling given by (\ref{weaklimit}), which are discarded by the observations, and the only influence of the kinetic coupling is on the values of the scalar field $\phi_{end}$ and $\phi_c$. For $n=5$, the scalar spectral index and the tensor-to-scalar ratio vary between the limits
$$
n_s^{(5)}=\frac{N-3}{N+2},\;\;\; r^{(5)}=\frac{32}{N+2},
$$
making it impossible to realize the slow-roll inflation between the current observed values of $n_s$ and $r$. For $n\ge 6$ it can be seen from (\ref{Eq:slow-roll-parameter-0}) that  $\phi_{end}$ and $\phi_i$ can be defined only for values of $\beta$ bellow some finite value that depends on $n$. Thus, taking $n=6$ for instance, the scalar field at the end of inflation becomes $\phi_{end}=\sqrt{8-\beta/2}$. This means that the strong coupling limit can not be defined and the slow-roll inflation takes the character of large-field inflation, making it impossible for the self coupling to take values in the interval $\lambda\ge 0.1$. 
Note also that as $n$ increases the spectral index moves out to the left of the Planck $95\%$ CL and starting from $n\ge 4$, $n_s$ is located outside that region, while $r$ moves to the top of the $95\%$ CL region,  being located outside the region for $n\ge 3$. In the opposite case, assuming $n<-1$, it is found that in the strong coupling limit the observables tend to the values
\be\label{stronglimit1}
n_s^{(n<-1)}=\frac{6N-n(N-1)-8}{N(6-n)+2},\;\; r^{(n<-1)}=\frac{32}{N(6-n)+2}.
\ee
This result favor $r$ that decreases as $n$ takes larger negative values, moving towards the $1\sigma$-region in Fig. 3, but the scalar spectral index increases moving outside the $2\sigma$-region to the right. Thus, for $n<-1$ it is not possible to reach values of $n_s$ consistent with the Planck 2018 observational data. In all  models considered above the observables $n_s$ and $r$ vary between the weak coupling limit, common to all cases, and the strong coupling limit for each case.\\


\section{Conclusions}
\label{section:conclusions}

\noindent We have presented a study of the slow-roll inflation driven by the Higgs Boson with non-minimal derivative coupling to curvature. The coupling function is of the power-law type and we have considered all possible scenarios with positive and negative power. 
\noindent The results depicted in Fig. \ref{rvsnskin} show that both tensor to scalar ratio $r$ and scalar spectral index $n_s$ depend on the parameter $\alpha=\lambda\beta$, where $\lambda$ is the self coupling of the Higgs potential and $\beta$ (the kinetic coupling parameter) is used to set the appropriate values of the scalar field at the horizon crossing. As can be inferred from Fig. \ref{phiini}, for large $\beta$, the rate of decreasing of the scalar field at the horizon crossing slows down as $\beta$ increases, and therefore the observables take practically constant values, very close to the strong coupling limit. The self coupling $\lambda$ is more sensitive to the variation of $\phi_i$ (and therefore of $\beta$) as shown in Figs. 3 and 4. \\
\noindent It was found that the only viable values for the power are $n=2,1,0,-1$, which may lead to observables that satisfy the latest \textit{Planck} constraints on $r$ and $n_s$ at the $95\% $ CL.  For large and small values of  $\beta$, that give correspondingly small and large values of the inflaton at the horizon crossing, it is possible to satisfy the COBE restrictions with adequate values of the self coupling constant $\lambda$ in a wide interval $\lambda\sim 10^{-11}-0.3$, that covers the Higgs boson phenomenology. Furthermore, the energy contribution of the kinetic term is small enough to be consistent with the slow-roll formalism, driven by the potential in the early stages of the inflation.\\
\noindent Considering all possible powers, it was shown that the models with $n\ge 3$ lead to tensor-to-scalar ratio values $r>0.1$, growing with the increase of $n$, while the scalar spectral index decreases, and starting from $n=4$ is located outside to the left of the $95 \%$ CL region depicted in Fig. 3. The analysis of power-law couplings with $n<-1$ shows that while $r$ is favored taking values well below the $r=0.1$ limit, the scalar spectral index increases considerably, moving outside the $95\%$ CL region to the right. Thus, for $n<-1$ it is not possible to reach values of $n_s$ consistent with the Planck 2018 observational data and these models are discarded. \\
According to the trajectories depicted in Fig. 1, the main flaw of the models is in the tensor-to-scalar ratio, where the only cases in which $r$ takes values below $0.1$ are $n=0$ and $n=-1$. One advantage of the models $n=2,1,0,-1$ is that the self coupling $\lambda$ can take values in a wide interval that covers different phenomenological limits.  
while  $n_s$ and $r$ (specifically for the cases $n=-1,0$) can be consistent with current observations. Particularly interesting is the model with constant coupling $\beta$ ($n=0$) which was also study quantitatively in \cite{Tsujikawa:2012mk,Yang:2015pga}  where it was also shown that the consistency with the Higgs boson is reached at large coupling regime.\\
Concerning the reheating process, a preliminary analysis for the present model shows that, when the kinetic coupling is given by an inverse power-law function, the reheating takes place in unusual form. Taking into account the absence of post-inflationary oscillations, the standard reheating mechanisms cannot successfully exit the inflation and reheat the universe. In other words, due to the fact that the coupling becomes very large at the end of inflation, strongly affecting the post-inflationary dynamics, then the simplest reheating process cannot take place. A possible solution to this problem is considering a coupling function of the form $F_1\sim (\phi+\phi_0)^{-n}$. Numerical calculations show that this modification allow the appearance of post-inflationary oscillations, leading to a viable reheating. We expect that this 
little displacement in the field has only an important effect at the end of inflation without substantially affecting the predictions for the power spectra. The study of  adequate restrictions for $\phi_0$ are being studied.\\
\noindent We have explored the possibilities for the scalar field model, with Higgs potential and derivative coupling to gravity, to achieve successful inflation, where the kinetic coupling function is a monomial of the scalar field. Although we have shown that the models with $n=2,1,0,-1$ are consistent with current observations by the \textit{Planck} collaboration (falling into the $95\% $ CL), they might be in trouble when including more data sets such as those by the BICEP2/Keck Array collaboration \cite{Array:2015xqh}. A more general kinetic coupling function (e.g., $F(\phi,X)$) or additional curvature corrections could be worked out in order to analyze whether or not a Higgs field with such curvature corrections can satisfy the most stringent constraints. 

\section*{Acknowledgments}
\noindent This work was supported by Universidad del Valle under project CI 71195.  WC and DFJ acknowledge financial support from COLCIENCIAS (Colombia).

\bibliographystyle{utcaps}
\bibliography{paper}

\end{document}